\documentclass[reprint,
               onecolumn,
               notitlepage,
               aps,
               pra,
               showemail,
               showpacs]{revtex4-1}

\usepackage{amsthm}
\usepackage{amsmath}
\usepackage{mathtools}
\usepackage{latexsym}
\usepackage{amsfonts}
\usepackage{amssymb}
\usepackage{color}
\usepackage{bbm,dsfont}
\usepackage{graphicx}
\usepackage{tikz}
\usepackage{placeins}
\usepackage[utf8]{inputenc}
\usepackage[pdfpagelabels=true]{hyperref}
\usepackage[normalem]{ulem}
\usepackage{setspace}
\usepackage[marginparwidth=3cm, marginparsep=2pt]{geometry}

\newcounter{mparcnt}

\newcommand{\real}{\mathbb R} %real
\newcommand{\hi}{\mathcal{H}} %Hilbert space H
\newcommand{\ket}[1]{|#1\rangle} %ket
\newcommand{\tr}[2]{\textrm{tr}_{#2}\left\{#1\right\}}
\newcommand{\id}{\mathbbm{1}} %identity operator
\newcommand{\transp}{^{\textrm{T}}}

\newcommand{\vR}{\mathbf{R}} %R
\newcommand{\vxi}{\boldsymbol{\xi}} %xi
\newcommand{\matt}[1]{\left( \begin{array}{cc} #1 \end{array} \right)} %2*2 matrix

\newcommand{\diff}{{\rm d}}

\begin{document}
\title{Typical Gaussian Quantum Information}

\author{Philipp Sohr}
\affiliation{Institut f{\"u}r Theoretische Physik, Technische Universit{\"a}t Dresden, 
D-01062,Dresden, Germany}

\author{Valentin Link}
\affiliation{Institut f{\"u}r Theoretische Physik, Technische Universit{\"a}t Dresden, 
D-01062,Dresden, Germany}

\author{Kimmo Luoma}
\email{kimmo.luoma@tu-dresden.de}
\affiliation{Institut f{\"u}r Theoretische Physik, Technische Universit{\"a}t Dresden, 
D-01062,Dresden, Germany}

\author{Walter Strunz}
\email{walter.strunz@tu-dresden.de}
\affiliation{Institut f{\"u}r Theoretische Physik, Technische Universit{\"a}t Dresden, 
D-01062,Dresden, Germany}

%\thanks{}


%\email{}

\date{\today}

\begin{abstract}
{We investigate different geometries and  
invariant measures on the space of mixed Gaussian
quantum states. We show that when the global purity of the 
state is held fixed, {these} measures coincide and 
it is possible, within this constraint, to define a unique {notion of volume} 
on the space of mixed Gaussian states.
We then use the so defined measure to study typical non-classical correlations
of two mode mixed Gaussian quantum states, in particular entanglement
and steerability. {We show that under the purity constraint alone, typical values
for symplectic invariants can be computed very elegantly, irrespectively of 
the non-compactness of the underlying state space. Then we consider 
finite volumes by constraining the purity and energy of the Gaussian
state and compute typical values of quantum correlations numerically.}}
\end{abstract}

%\keywords{}
\maketitle

\section{Introduction}\label{sec:intro}
{Typical properties are interesting for example,
from the viewpoint of the emergence of thermodynamic behavior of 
{many-particle} quantum systems~\cite{gemmer2009quantum} and 
{for} the study of entanglement~\cite{Hayden2006}, {the} latter being  
also in the focus of this article. There is hope that the study of 
quantum correlations in mixed multipartite systems can be greatly simplified
by looking at the properties of typical states~\cite{Dahlsten2014}.}

{Typical here means that we consider a uniform distribution in the space of all states.}
{A geometry on the state space has
to be fixed, which then gives rise to a unitarily invariant volume element.  
For pure states there is a unique unbiased measure which emerges from
the Haar measure on the unitary group~\cite{Lubkin1978}. For mixed quantum states 
no such unique measure exists~\cite{Bengtsson2006}.
{Investigations of typical properties of mixed states of  quantum systems
were first pioneered in~\cite{Zyczkowski1998,Zyczkowski1999}. Most results so far
are for systems with finite dimensional Hilbert space.}

{{We will focus our investigations on the geometry and typical properties
of mixed Gaussian states, which form a subspace of  
continuous variable quantum states.} Gaussian states are important for two main reasons, they can 
be created and manipulated experimentally using linear optics~\cite{Ferraro2005,Adesso2014} 
and they are completely characterized {by a finite number of parameters, the first and the second 
moments of canonical position- and momentum operators \cite{holevo2011probabilistic}}.
{Gaussian states are best represented by their (positive) Gaussian Wigner function~\cite{Adesso2007}.}}

{For pure Gaussian states, a {unique unbiased} measure using the {invariant} Haar measure on 
the symplectic group has been constructed in~\cite{Lupo2012}. 
{For mixed Gaussian states, {different invariant} measures have been constructed in~\cite{Link2015} using
the Hilbert-Schmidt metric and in~\cite{Felice2017} using ideas 
from information geometry}. Even though Gaussian states are easy to
characterize, the state space is not compact, which is
related to the possibility of having arbitrarily squeezed states~\cite{Lupo2012}.  
Therefore, the constructed measures are not normalizable unless some 
further restrictions are made such as fixing the energy of the state~\cite{Serafini2007,Lupo2012}.} 

\begin{figure}[!h]
  \begin{tabular}{lr}
    \includegraphics[width=0.5\textwidth]{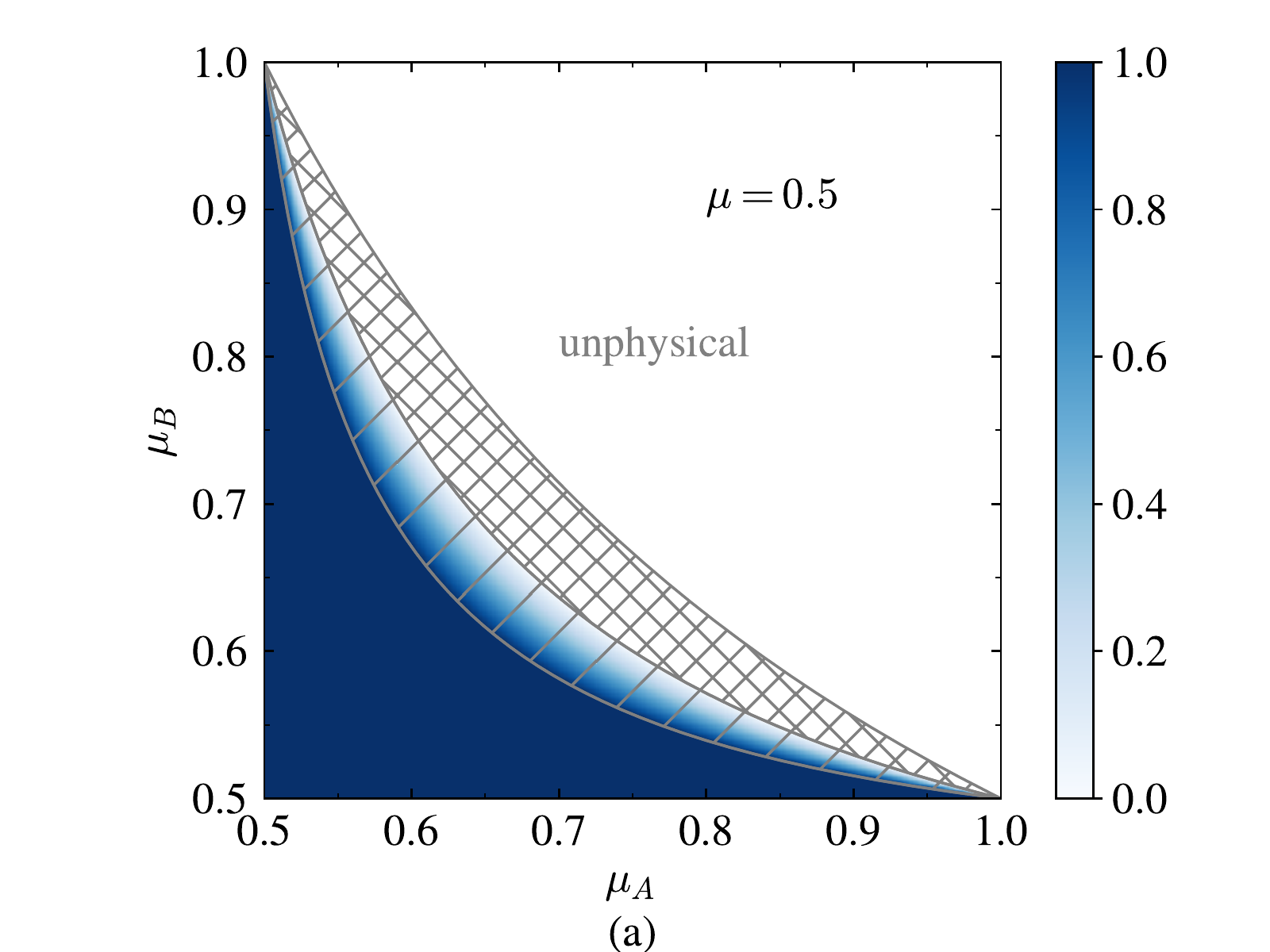} 
    &
      \includegraphics[width=0.5\textwidth]{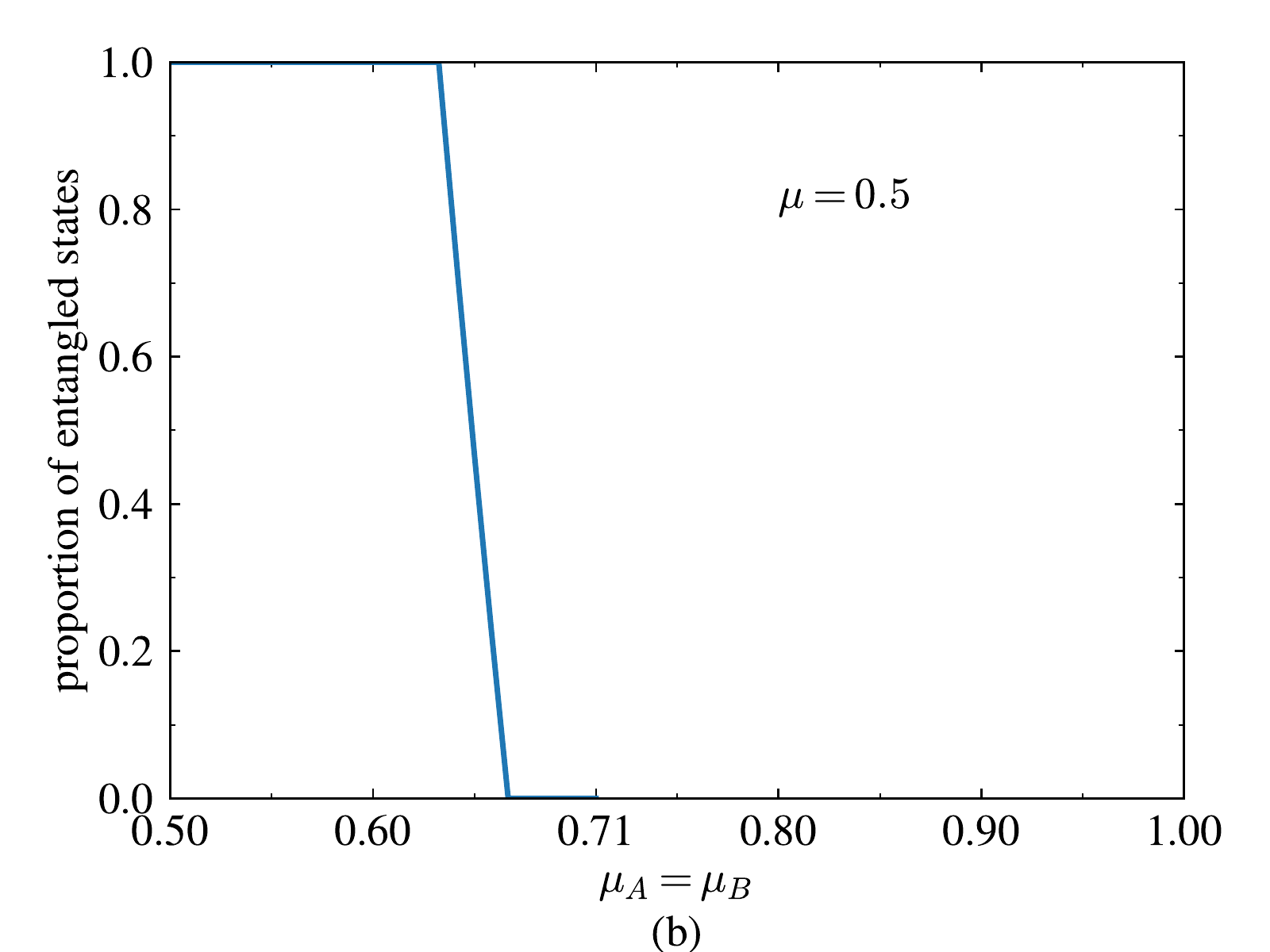}
  \end{tabular}  
  \caption{a) Domain of separable states (double hatched),
    entangled states (dark blue) and coexistence of entangled and separable states (single hatched) for two mode Gaussian states. 
    The shading indicates the proportion of
    entangled states as a function of marginal purities $\mu_A$ and $\mu_B$
    when the global purity $\mu = 0.5$. In the unhatched white region no physical states exist.
    b) Proportion of entangled states in the physical domain when $\mu_A=\mu_B$ and $\mu=0.5$.}%  We see that 
    % for purities $0.5 \leq \mu_{A/B} \lessapprox 0.63$ all states are entangled and 
    % the fraction of entangled decreases linearly when $0.63\lessapprox \mu_{A/B} \lessapprox 0.66$.
    % When $0.66\lessapprox \mu_{A/B}\lessapprox 0.71 $ all states are separable and no physical
    %states exist for which $\mu_{A/B}\gtrapprox 0.71$}
  \label{fig:regionzoom}
\end{figure}

{Our main results are the following. {We study three 
very differently motivated measures on mixed Gaussian states and we can show analytically that when
the purity of the Gaussian state is fixed, these three measures are equivalent up to a constant.} 
This type of phenomenon was first
observed numerically in~\cite{Zyczkowski1999} for finite dimensional mixed quantum
states. 
The observation allows us to propose a unique measure for mixed Gaussian states with fixed purity. 
{Finally}, we apply our results in order to study typical quantum correlations in
bipartite Gaussian states. 
{As an example we show in {Fig}.~\ref{fig:regionzoom} a)
the proportion of entangled two-mode Gaussian states 
as a function of local purities $\mu_A$, $\mu_B$ for a fixed global purity $\mu$. 
The single hatched area in the figure was identified in \cite{Adesso2004} to be a so called 
coexistence region, where it is not possible {to} discriminate whether a state is 
entangled or separable by purity measurements alone. 
In Fig.~\ref{fig:regionzoom} b) the proportion of entangled states along a cut $\mu_A=\mu_B$ trough the physical domain is shown while
the global purity $\mu=0.5$. 
Interestingly we see that in the coexistence region the 
propotion of entangled states decreases linearly with $\mu_{A/B}$ 
and reaches zero in the separable region. The boundary 
of the physical domain is at $\mu_{A/B}=\sqrt{\mu}\approx 0.71$.
% Interestingly, we see that 
% for marginal purities $0 \leq \mu_{A/B} \lessapprox 0.63$ all states are entangled, while
% the portion of entangled decreases linearly when $0.63\lessapprox \mu_{A/B} \lessapprox 0.66$.
% When $0.66\lessapprox \mu_{A/B}\lessapprox 0.71 $ all states are separable and no physical
% states exist for which $\mu_{A/B}\gtrapprox 0.71$.
With a measure at hand we can even go beyond the results presented in Fig.~\ref{fig:regionzoom} and 
characterize also the \emph{amount} of typical entanglement.}

{The outline of the article is the following. In Sec. \ref{sec:cv-systems} we review the main properties of 
Gaussian states and introduce symplectic invariants that are relevant
for {this} article. In Sec. \ref{sec:invar-meas-mixed} we construct
three different measures for mixed Gaussian states and show that they are 
equivalent when global purity is fixed. In Sec. \ref{sec:typic-quant-corr} we 
{review} different classes of quantum correlations such as 
entanglement and steerability and their quantitative
measures for Gaussian states. In Sec.~\ref{sec:comp-typic-corr} we
discuss typical quantum correlations of two mode Gaussian states. To
deal with the issue of non-compactness we apply two different
strategies. First the typical entanglement as a function of marginal
purities and with fixed global purity, as in Fig.~\ref{fig:regionzoom}, is discussed. 
Then we examine compact subspaces
of mixed Gaussian states with fixed global purity given by
constraining the energy of the states. Lastly, in Sec. \ref{sec:concl}
we conclude. We have collected many of the technical details and
computations in
Appendices~\ref{sec:deriv_FR},~\ref{sec:two-mode-hilbert}
and~\ref{sec:deriv_Econstraint}.}

\section{Continuous Variable Systems and Gaussian Quantum States}\label{sec:cv-systems}
{A continuous variable (CV) system is a quantum system with
Hilbert space  $\hi_i$ isomorphic to $L^2(\real)$. The description of an $N$-mode CV
system is based on the Hilbert space $\hi={\bigotimes\limits_{i=1}^N} \hi_i$. In this space the canonical position- and momentum operators 
\begin{align}\label{eq:creation_annihilation}
  \hat q_i=\hat a_i+\hat a_i^\dagger\,, && \hat p_i=i(\hat a_i^\dagger -\hat a_i),
\end{align}
are related to the creation and annihilation operators acting on Fock states in the usual manner, with commutation relations 
$[\hat a_i,\hat a_j^\dagger]=\delta_{ij}$ and 
$[\hat a_i,\hat a_j]=0=[\hat a_i^\dagger,\hat a_j^\dagger]$. {We follow the
standard conventions in this field leading to a commutation relation
$[\hat q_k,\hat p_l]=2i\delta_{kl}$.} For notational convenience we group together the canonical operators in a single vector
\begin{align}
 \hat\vR=(\hat q_1,\hat p_1,...,\hat q_N,\hat p_N)\,,\qquad  [\hat R_i,\hat R_j]=2\mathrm{i}\Omega_{ij}\,,\qquad \Omega=&\bigoplus_{i=1}^N\matt{0 & 1 \\ -1 & 0},
\end{align}
where $\Omega$ is called the symplectic matrix. A CV quantum state given by a density matrix $\hat\rho$ can be represented in phase space by the Wigner function
\begin{align}\label{eq:Wigner_transf}
  W(\boldsymbol{x})=&\int_{\real^{2n}}\frac{\diff^{ 2n}\vxi}{2\pi^{2n}}e^{-i\boldsymbol{x}\transp \Omega\vxi}
           \tr{\hat\rho \hat D(\vxi)}{},\qquad
  \hat D(\vxi)=e^{i\hat\vR\transp \Omega\vxi},
\end{align}
with $\boldsymbol{x}\in \real^{2N}$ and $\hat D(\vxi)$ is the Weyl- or shift operator. A state is called Gaussian {iff} it has a Gaussian Wigner function
\begin{align}\label{eq:Gaussian_state}
  W(\boldsymbol{x})=\frac{1}{\pi\sqrt{\mathrm{det}\Sigma}}\exp\left(-\frac{1}{2}(\boldsymbol{x}-\boldsymbol{l})\transp \Sigma^{-1}(\boldsymbol{x}-\boldsymbol{l})\right).
\end{align}
The quantities $l_i= \tr{\hat\rho\hat R_i}{}$ and 
$\Sigma_{ij}=\frac{1}{2}\tr{\hat\rho\left(\hat R_i\hat R_j+\hat R_i \hat R_j\right)}{}-l_i l_j$
are called the displacement vector and the covariance matrix, respectively. The displacement
of any Gaussian state can be brought to zero by action of {a} local shift operator. 
Therefore $\boldsymbol{l}$ contains no information about the correlations between modes and will 
be set to zero in the following. An $N$-mode Gaussian state is then fully characterized by its covariance matrix (CM). In fact, any real symmetric matrix $\Sigma$ that satisfies
\begin{align}\label{eq:bonafide}
  \Sigma + i\Omega \geq 0,
\end{align}
is a CM of a Gaussian quantum state.
Unitary operations $U_G$ generated by self-adjoint operators that are 
quadratic in the canonical operators preserve Gaussianity of a state. If $\Sigma$ and $\boldsymbol{l}$ are 
the covariance matrix and displacement vector of the Gaussian state $\hat\rho$ and 
$\hat\rho\mapsto\hat\rho'=\hat U_G\hat\rho \hat U_G^\dagger$, then
\begin{align}\label{eq:Gaussian_transf}
  \Sigma \mapsto \Sigma'=S\Sigma S\transp \,,\qquad \boldsymbol{l}\mapsto\boldsymbol{l}'=S\boldsymbol{l}\,,
\end{align}
where $S \in \mathrm{Sp}(2N)$ is a symplectic transformation $S\transp \Omega S=\Omega$ \cite{Simon1988}.
Gaussian unitaries can be used to diagonalize a Gaussian density operator: For any covariance matrix $\Sigma$ there exists a symplectic
transformation $S$ such that $S\transp \Sigma S$ is diagonal, with each
diagonal entry appearing twice (Williamson form \citep{Williamson1936})
\begin{align} S\transp \Sigma S = D
=\bigoplus_{i=1}^N\begin{pmatrix}
                   \nu_i &0\\
                   0 & \nu_i
                  \end{pmatrix}
\,,\qquad \nu_i\geq
1\,.
\label{eq:D}
\end{align}
The $N$ values $\nu_i$ are called the symplectic
eigenvalues of $\Sigma$. They characterize a Gaussian state up to
unitary transformations, and are thus equivalent to the
eigenvalues of a density operator $\hat \rho$. The density operator corresponding
to {a} diagonal covariance matrix $D$ is a tensor product of thermal
states.  This state is pure if and only if all $\nu_i$ are equal to one, that is if it is the
vacuum state.
The purity of a Gaussian quantum state is given by the inverse of the product of the symplectic eigenvalues
\begin{equation}
 \mu(\Sigma)=\frac{1}{\sqrt{\det \Sigma}}=\prod_{k=1}^N\frac{1}{\nu_k}\,.
\end{equation}
As for a symplectic matrix $\det S = 1$, the purity is invariant under symplectic transformations of $\Sigma$. Another symplectic invariant relevant for this article is the \emph{seralian} \citep{Adesso2004}
\begin{equation}
\Delta(\Sigma) = \sum_{k=1}^N \nu_k^2 
%= \sum_{i=1}^N \det\sigma_{i,i}+\2\sum_{j<i=1}^N \det \sigma_{j,i}
\, .
\end{equation}
If the total state of a $N$-mode CV system composed of system $A$ and $B$ 
consisting of $N_A$ and $N_B=N-N_A$ modes is Gaussian, 
then also the reduced states are Gaussian and the corresponding CMs $\Sigma_A$ and $\Sigma_B$ are two diagonal blocks in the total CM
\begin{equation}
 \Sigma=\begin{pmatrix}
         \Sigma_A &C\\
         C\transp  & \Sigma_B
        \end{pmatrix}\,.
\end{equation}
The marginal purities $\mu_A=\frac{1}{\sqrt{\det \Sigma_A}}$ and 
$\mu_B=\frac{1}{\sqrt{\det \Sigma_B}}$ are the purities of the reduced states. A symplectic transformation is called local {if} $S=S_A\oplus S_B$ with $S_A\in\mathrm{Sp}(2N_A),\,S_B\in\mathrm{Sp}(2N_B)$. These transformations correspond to local unitary operations and do not change non-local correlations between $A$ and $B$ such as entanglement. \\
{Following \cite{Lupo2012}, we also define the ``energy''} of a CV state with respect to the Hamiltonian
\begin{align}\label{eq:Hamiltonian} \hat H =\sum_i\frac{1}{2}(\hat q_i^2+\hat p_i^2).
\end{align}
Its expectation value for a Gaussian state is
\begin{align}\label{eq:Energy} E=\frac{1}{2}&\tr{\hat\rho\hat
H}{}=\frac{1}{2} \tr{\Sigma}{}+\frac{1}{2}\boldsymbol{l}\transp \boldsymbol{l}.
\end{align}
{One should rather think of $E$ being proportional to the number
of excitations in the state $\hat\rho$.}}

\section{Invariant Measures for mixed Gaussian states}\label{sec:invar-meas-mixed}
{In the space of pure quantum states there exists a unique notion of
volume given by the invariant measure of the unitary group. In
particular any pure quantum state can be written as a unitary
transformation of a fixed pure state $\ket{\psi}=U\ket{\psi_0}$. Thus the
invariant measure (Haar measure) on the unitary group gives a
non-biased measure for pure quantum states \cite{Bengtsson2006}. The same holds of course
for pure Gaussian states: Since the covariance matrix of a pure
Gaussian state can be written as $\Sigma=S\transp S$, the invariant measure
on the symplectic group $\text{Sp}(2N)$ is a unique measure in the space of pure
Gaussian quantum states. This has been studied in
\cite{Lupo2012}.\\
Considering mixed states, there no longer exists a unique invariant
measure, since there is the additional non-unitary freedom in the
eigenvalues of the density matrix \cite{Hall1998}, or equivalently the symplectic
eigenvalues in the case of Gaussian states. The volume element of any
invariant measure for mixed Gaussian states can then be written as \cite{Link2015}
\begin{equation}
 \begin{gathered} \text{d}V=P(\nu_1,...,\nu_N)\,\text{d}\mu_N
(S)\prod\limits_{i=1}^{N}\text{d}\nu_i
 \end{gathered}
\end{equation} where $\text{d}\mu_N (S)$ denotes the invariant measure
on the symplectic group $\text{Sp}(2N)$ and $P$ is a probability
density of the eigenvalues $\nu_i$. In the following we compare three
very differently motivated measures by exploiting this decomposition.}
\subsection{Comparison of invariant measures}
\subparagraph{Hilbert-Schmidt}
The Hilbert-Schmidt measure is a natural {and easily computable measure on the 
space of operators acting on $\hi$}
induced by the unitarily invariant Hilbert-Schmidt metric
$\text{d}s^2_{HS}=\text{tr}(\text{d}\rho^{\,2})$. Confining this to
the manifold of Gaussian quantum states gives a metric on the space of
admissible covariance matrices $\Sigma$
\cite{Link2015}
\begin{equation}
\text{d}s_{\text{HS}}^2=\frac{1}{16\sqrt{\det\Sigma}}\Big(\big(\text{tr}(\Sigma^{-1}\text{d}
\Sigma)\big)^2+2\text{tr}\big((\Sigma^{-1}\text{d}
\Sigma)^2\big)\Big)\,.
 \label{eq:HS_dist}
\end{equation} Expressing covariance matrices by the symplectic
eigenvalue decomposition $\Sigma=S\transp  D S$ the volume element of the
induced measure can, up to a constant, be written as
\begin{align}\label{eq:HS_measure}
   \text{d}V_{\text{HS}}&=P_{\text{HS}}(\nu_1,...,\nu_N)\,\text{d}\mu_N
   (S)\prod\limits_{i=1}^{N}\text{d}\nu_i\,, \\
  \label{eq:HS_pdensity}
   P_{\text{HS}}(\nu_1,...,\nu_N)&=\Big(\prod_{k=1}^N\nu_k\Big)^{-N(N+\frac{5}{2})+1}
   \prod\limits_{l>m=1}^N(\nu_l^2-\nu_m^2)^2\,.
\end{align}
{Detailed computations and the derivation of Eqs.~(\ref{eq:HS_measure}) and (\ref{eq:HS_pdensity}) can be found in \cite{Link2015}.}
\subparagraph{Fisher-Rao}
{The fact that Gaussian CV quantum states have a positive Wigner function everywhere allows to borrow 
ideas from classical information geometry and apply them, at least formally, to 
quantum systems in the Gaussian domain. One such idea is to use Fisher-Rao metric, which is a metric in 
the space of probability distributions~\cite{amari1990differential}, 
as a metric for the space of Gaussian quantum states.}
{We would like to stress that Gaussian quantum states are not classical states, and the similarity
of formalism between classical and quantum phase space distributions should not be pushed too far, since
quantum mechanics is a fundamentally non-commutative theory.}

{Following
\cite{Felice2017}, the Fisher-Rao metric} can be expressed with covariance matrices as
\begin{equation}
\text{d}s_{\text{FR}}^2=\frac{1}{2}\text{tr}\big((\Sigma^{-1}\text{d}
\Sigma)^2\big)\,.
\end{equation} Note that a similar term appears also in Eq.~(\ref{eq:HS_dist}). {Thus to express the measure in terms of symplectic
eigenvalues, large parts of the results from \cite{Link2015} for the 
derivation of the Hilbert-Schmidt measure can be utilized.} The resulting expression is 
\begin{align}
  \label{eq:FR_measure}
  \text{d}V_{\text{FR}}&=P_{\text{FR}}(\nu_1,...,\nu_N)\,\text{d}\mu_N
                         (S)\prod\limits_{i=1}^{N}\text{d}\nu_i\,, \\
  \label{eq:FR_density}
  P_{\text{FR}}(\nu_1,...,\nu_N)&=\Big(\prod_{k=1}^N{\nu_k}\Big)^{-2N+1}
                                  \prod\limits_{l>m=1}^N(\nu_l^2-\nu_m^2)^2\,\,.
\end{align}
{To be more self contained, we have included the derivation of Eqs.~(\ref{eq:FR_measure}) and (\ref{eq:FR_density})
in Appendix \ref{sec:deriv_FR}.}
\subparagraph{Reduced states of pure Gaussian states}
A practical scheme to sample $N$-dimensional mixed states from the Hilbert-Schmidt measure is 
to sample $N^2$-dimensional pure states from the Haar measure  and partially trace over $N$ degrees of freedom. This method
has been applied to qubits and finite dimensional systems, see for example~\cite{Bengtsson2006,Milz2014}.
{Here, we consider sampling pure Gaussian states from the Haar measure, introduced in \cite{Lupo2012}, 
with doubled mode number $2N$ and partially trace out $N$ modes. In order to obtain a useful representation for this measure, we write the CV of the
$2N$ mode pure state in the form
${\Sigma}={S}\transp \sigma {S}$, with ${S}=S_A\oplus
S_B$, where $S_A,S_B\in\mathrm{Sp(2N)}$ and \cite{Adesso2007}
\begin{equation}
\sigma=\begin{pmatrix} D & C\\ C & D
        \end{pmatrix}\,,\quad
D=\bigoplus_{i=1}^N\begin{pmatrix}
                               \nu_i &0\\
                               0& \nu_i
                              \end{pmatrix}\,,\quad
C=\bigoplus_{i=1}^N\begin{pmatrix}
                    \sqrt{\nu_i^2-1} &0\\
                    0& -\sqrt{\nu_i^2-1}
                   \end{pmatrix}\,.
\end{equation} 
{The symplectic eigenvalues {of} the $N$-mode subsystems $A$ and $B$
are denoted with $\nu_i$. The volume element can be found in \cite{Lupo2012} 
and it reads
\begin{equation}
 \begin{gathered} \text{d}\mu_{2N}
(S)=P_{2N}(\nu_1,...,\nu_N)\,\text{d}\mu_N (S_A)\,\text{d}\mu_N
(S_B)\prod\limits_{i=1}^{N}\text{d}\nu_i\,, \\
P_{2N}(\nu_1,...,\nu_N)=\Big(\prod_{k=1}^N{\nu_k}\Big)^{2}
\prod\limits_{l>m=1}^N(\nu_l^2-\nu_m^2)^2\,\,.
 \end{gathered}
\end{equation} 
Tracing out $N$ modes then just corresponds to integrating over one of the local symplectic groups giving rise merely to a constant factor. 
The density of symplectic eigenvalues $P_{2N}$ then defines an
invariant measure for mixed $N$-mode Gaussian states.}

\subsection{Unique fixed purity measure}
We observe that the probability densities  $P_{HS},\, P_{FR}$ and $P_{2N}$  
over the symplectic eigenvalues $\nu_1,\ldots,\nu_N$  only differ by a prefactor, which is a
power of the purity. Therefore, if we consider Gaussian states with a fixed
purity then all of the three  measures are identical up to a constant. 
Since all three measures are invariant measures this statement is
trivial for $N=1$ and pure states $\mu=1$. This strongly motivates us 
to consider Gaussian states of fixed purity and the measure
\begin{equation}
 \begin{gathered}
 \text{d}V_{\mu}=P_{\mu}(\nu_1,...,\nu_N)\,\text{d}\mu_N (S)\prod\limits_{i=1}^{N}\text{d}\nu_i\,, \\ 
 P_{\mu}(\nu_1,...,\nu_N)=\delta\Big(\mu-\prod_{k=1}^N\frac{1}{\nu_k}\Big)\prod\limits_{l>m=1}^N(\nu_l^2-\nu_m^2)^2\,\,.
 \end{gathered}
 \label{eq:fixed_purity_measure}
\end{equation}
We have shown that when the global purity is fixed, three very 
different measures on the set of mixed Gaussian states are equivalent up 
to a constant. Similar studies for finite dimensional systems were done in 
\cite{Zyczkowski1999}. There, the numerical data sampled from different invariant 
measures conditioned on purity showed close but not perfect agreement, 
{in contrast} to our analytical findings for Gaussian states.
\section{Entanglement and Steerability of two mode Gaussian states\label{sec:typic-quant-corr}}
The proposed volume element (\ref{eq:fixed_purity_measure}) allows us to study 
in detail the typical correlation properties of two mode Gaussian states. In this 
section necessary measures to quantify entanglement and also quantum steering are introduced. 
Then, in Sec. \ref{sec:comp-typic-corr} we compute the typical values of such quantum correlations.
\subparagraph{Entanglement}
For $1\times N$-mode Gaussian states the 
positive partial transpose (PPT) or
Peres-Horodecki criterion, introduced for CV systems by Simon
\citep{Simon2000} is necessary and sufficient condition 
for separability.
The transpose of a density matrix $\hat\rho$
corresponds to a mirror reflection in phase space, which means that
the sign of the momentum flips. The non-unitary partial transpose
operation corresponds to {an} inversion of the momenta of one party
only. For the $1\times 1$-mode Gaussian case with the partial
transpose applied to the second mode, this is written as $\vR \mapsto
\Lambda \vR = \tilde{\vR} = (q_1,p_1,q_2,-p_2)\transp $ with $\Lambda
= \textrm{diag}(1,1,1,-1)$. 
A Gaussian state with CM
$\Sigma$ is separable iff the CM after partial transposition satisfies
the bona fide condition \eqref{eq:bonafide} i.e. is the CM of a
physical state.

The logarithmic negativity
\begin{equation}
E_N = \textrm{max}\left\lbrace 0, -\textrm{log}_2 \tilde{\nu}_-\right\rbrace\,,
\label{eq:logarithmicnegativity}
\end{equation} 
with $ 2 \tilde{\nu}^2_\pm = \tilde{\Delta} \pm
\sqrt{\tilde{\Delta}^2-4/ \mu^2}$ and $ \tilde{\Delta} = 2/\mu_A +
2/\mu_B - \Delta$ is a quantitative measure for entanglement as it measures the 
degree of violation of the PPT
criterion~\cite{Plenio2005}.
\subparagraph{Steerability}
In the hierarchy of quantum correlations, 
steering is a distinct class for general quantum states. It is stronger than entanglement but weaker then non-locality, 
and it is inherently asymmetric~\cite{Wiseman2007,Jones2007}. 
In a typical steering scenario there {are} two parties, Alice and Bob, who
share a quantum state. Alice has some fixed
set of measurements, described by a set of positive operator
valued measures that she can perform locally and Bob
can do local state tomography.  If the state is $A\to B$ steerable, Alice can then, by measuring her local 
{observables}, steer Bobs state to such
state assemblages that they cannot be described by any local hidden state
model. Similarly, the state is $B\to A$ steerable if the roles of 
Bob and Alice are interchanged.
 
%In the realm of Gaussian states and Gaussian measurements,
%a bipartite state with covariance matrix $\Sigma_{AB}$ is $A\to B$ steerable iff \cite{Wiseman2007,Kogias2015}
%\begin{align}\label{eq:steer_cond}
%  \Sigma_{AB}+\vnull\oplus i\Omega_B < 0 
%\end{align}
%where $\vnull$ is a $2\times 2$ null matrix and 
%$\Omega_B$ is a symplectic matrix for Bob's mode. 
An operational criterion for {Gaussian} $A \to B$ steering is comparable to the bona fide condition {in Eq.}~\eqref{eq:bonafide}, 
{requires} only Gaussian measurements to be made, {and leads to a feasible expression in terms 
of the covariance matrix of the joint state~\cite{Wiseman2007}.} A $1\times 1$-mode Gaussian state is 
$A\to B$ {steerable} iff $\mu > \mu_A$ is satisfied~\cite{Kogias2015}.
%Further for $1\times 1$-mode Gaussians, the condition 
%(\ref{eq:steer_cond}) is equivalent to $\mu > \mu_A$~\cite{Kogias2015}.
Again, 
by interchanging the roles of Alice and Bob, the criterion 
for $B\to A$ steerability is obtained.
We call a state steerable if it is
$A\to B$ or $B\to A$  steerable.

A  quantitative measure for $A\to B$ steering is~\cite{Kogias2015}
\begin{equation}\label{eq:steering_measure}
G^{A\rightarrow B}(\Sigma) = \textrm{max}\left\lbrace 0,
 \textrm{ln} \left( \frac{\mu}{\mu_A}\right) \right\rbrace\,.
\end{equation}
$G^{A\rightarrow B}$ as well as $G^{B\rightarrow A}$ will never exceed the logarithmic negativity \eqref{eq:logarithmicnegativity}. The 
steering measure (\ref{eq:steering_measure}) quantifies the violation 
of the $1\times 1$-mode Gaussian $A\to B$ steering criteria. 
We call the maximum of $G^{A\rightarrow B}(\Sigma)$ and $G^{B\rightarrow A}(\Sigma)$ the steerability
\begin{equation}
G(\Sigma)=\mathrm{max}\left\lbrace 0, \mathrm{ln} \left( \frac{\mu}{\mu_A}\right), \mathrm{ln} \left( \frac{\mu}{\mu_B}\right) \right\rbrace\,.
\label{eq:symmetricsteerability}
\end{equation}

\section{Typical quantum correlations of two mode Gaussian states\label{sec:comp-typic-corr}}

As we have mentioned, the three measures will provide equivalent
statistical information for fixed purity. Without loss of generality
we choose to construct the volume element from the Hilbert-Schmidt
measure \eqref{eq:HSmuVolumeelement}.  Up to local symplectic
transformations, irrelevant for non-local correlations, any {two}
mode Gaussian state is completely characterized by the purity $\mu$ of
the state, the two marginal purities $\mu_A$, $\mu_B$ of the one mode
subsystems, and the seralian $\Delta$ \cite{Adesso2007,Adesso2014}. It
is therefore advantageous to express the Hilbert-Schmidt volume
element in these variables. Following the calculations in appendix
\ref{sec:two-mode-hilbert} we find the simple expression
\begin{align}
  \diff V_{\mathrm{HS}}=& \frac{\sqrt{3}}{512}\frac{\mu^7}{\mu_A^3\mu_B^3}\diff\mu_A\,\diff\mu_B\,\diff\mu\,\diff\Delta\,\diff\mu_1(S_A)\diff\mu_1(S_B),
  \label{eq:HSmuVolumeelement}
\end{align}
where $\diff\mu_1(S_{A})\diff\mu_1(S_B)$ is the invariant measure of the \textit{local} symplectic transformations. Even after constraining the purity to a fixed value when computing the volume of Gaussian states there appear two divergences related to arbitrarily strong squeezing. Firstly, the volume of the non-compact local symplectic group is infinite due to single-mode squeezing. Secondly, unbounded two-mode squeezing allows for arbitrarily small marginal purities. To circumvent this problem, further restrictions on the states considered have to be made. We propose two different strategies: Either fixing the marginal purities or fixing the energy of the states.

\subsection{Purity constrained typical quantum correlations}
The global and marginal purities of Gaussian states are easily
accessible in experiments and knowledge thereof may already be
sufficient to decide whether a mixed Gaussian state is entangled or
not \cite{Adesso2007,Adesso2014}. With a measure at hand we can now
quantify the typical \textit{amount} of entanglement expected for
given purities. This way we can also characterize a region where
purity measurements alone cannot determine if a state is entangled or
not, further referred to as the coexistence region.  In particular,
the typical value of, for instance, the logarithmic negativity $E_N$ for
Gaussian states of fixed purities $\mu, \mu_A, \mu_B$ is given by
\begin{align}
 \left\langle E_N\right\rangle |_{\mu,\mu_A,\mu_B} &= \frac{\int_{\Sigma+\mathrm{i}\Omega\geq 0}\mathrm{d}\Delta\diff\mu_1(S_{A})\diff\mu_1(S_B) E_N(\mu, \mu_A, \mu_B, \Delta)}{\int_{\Sigma+\mathrm{i}\Omega\geq 0}\mathrm{d}\Delta\diff\mu_1(S_{A})\diff\mu_1(S_B)}\notag \\ &=
 \frac{\int_{\Delta_{min}}^{\Delta_{max}}\mathrm{d}\Delta E_N(\mu, \mu_A, \mu_B, \Delta)}{\Delta_{max}-\Delta_{min}}\,.
\label{eq:expectationvalDelta}
\end{align}
The bona fide condition ${\Sigma+\mathrm{i}\Omega\geq 0}$ indicates
that the domain of integration is the set of admissible covariance
matrices belonging to physical Gaussian states. Note that divergent
contributions corresponding to the volume of the local symplectic
group cancel because $E_N$ is a local symplectic
invariant. {Thus, even though the volumes of the non-compact
subspaces are infinite, their ratio is still finite so that 
typical values such as (\ref{eq:expectationvalDelta}) can be computed.} $\Delta_{min/max}$ are the limits for the seralian
resulting from the bona fide condition. The integral can be solved
analytically. In Fig. \ref{fig:regions} we provide the results for
three different values of $\mu$. Naturally high purities and small
marginal purities correspond to a large amount of entanglement. In the
double hatched region only separable states exist and thus the typical
value of $E_N$ is zero. The single hatched area is the before mentioned
coexistence region, and the shaded area without hatching corresponds
to values of marginal purities where all states are entangled. For
marginal purity values in the white and unhatched area no physical
states exist. 
{In Fig.~\ref{fig:cut_trough_regions} we show a cut along
$\mu_A=\mu_B$ trough the domain of physical states for 
three different values of $\mu$. In each case
the the average entanglement decreases in non-linear fashion
for increasing marginal purity. As the marginal
purity increases, average entanglement 
decreases to zero in the separable domain and ceases to 
be well defined for marginal purities that are in the 
unphysical domain $\mu_{A/B}>\sqrt{\mu}$. We stress that if we would plot 
again the proportion of entangled states for a fixed  
global purity as in Fig.~\ref{fig:regionzoom} b)
we would see linear behavior for any linear cut 
trough the coexistence region.}

{We do not investigate typical 
steering under the purity constraint, since the local and global purities alone 
already completely determine the steerability of a Gaussian state~\cite{Kogias2015}.}

\begin{figure}[!t]
\includegraphics[width=\textwidth]{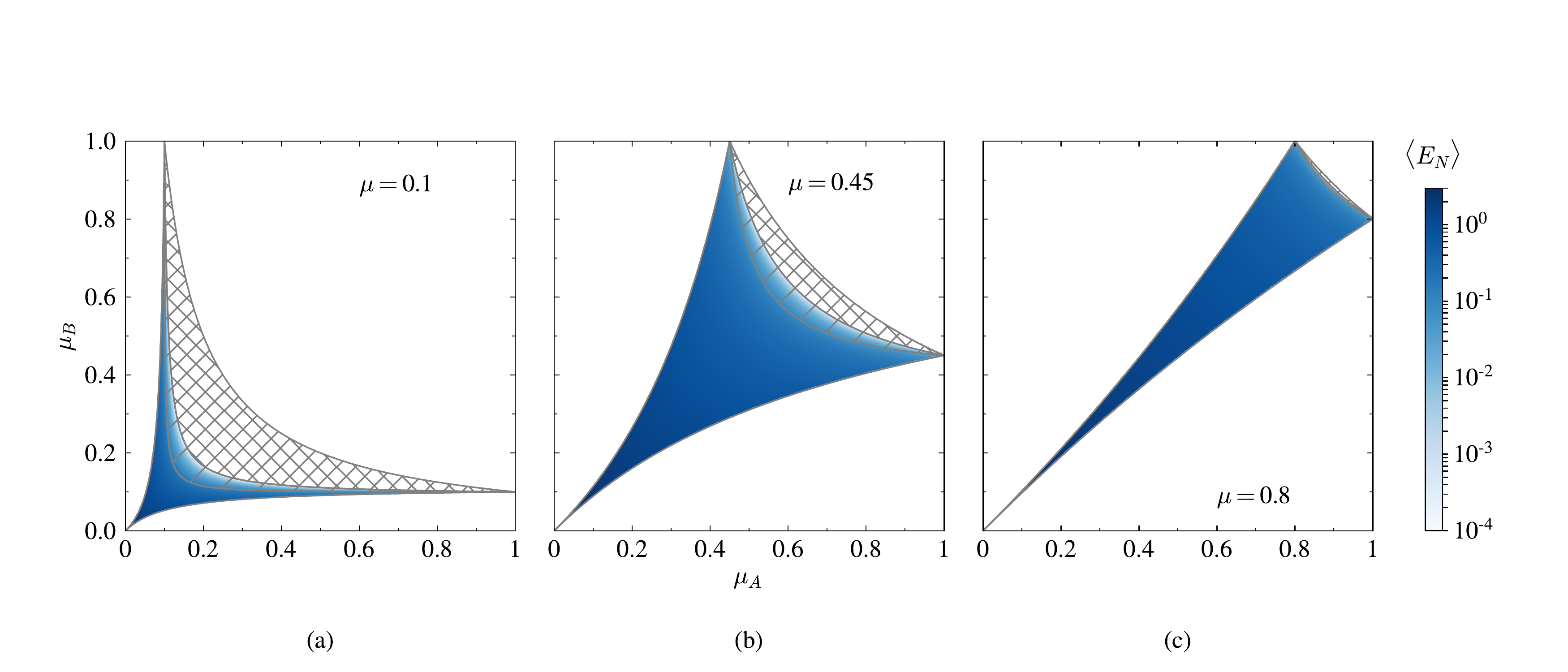}
\caption{Regions of separable states (double hatched), coexistence 
(single hatched and shaded) and entangled states (shading only) for
fixed marginal purities and global purities 0.1, 0.45 and 0.8. The
shading indicates the typical logarithmic negativity. In the unhatched
white region no physical states exist.}
\label{fig:regions}
\end{figure}

\begin{figure}[!h]
\includegraphics[width=\textwidth]{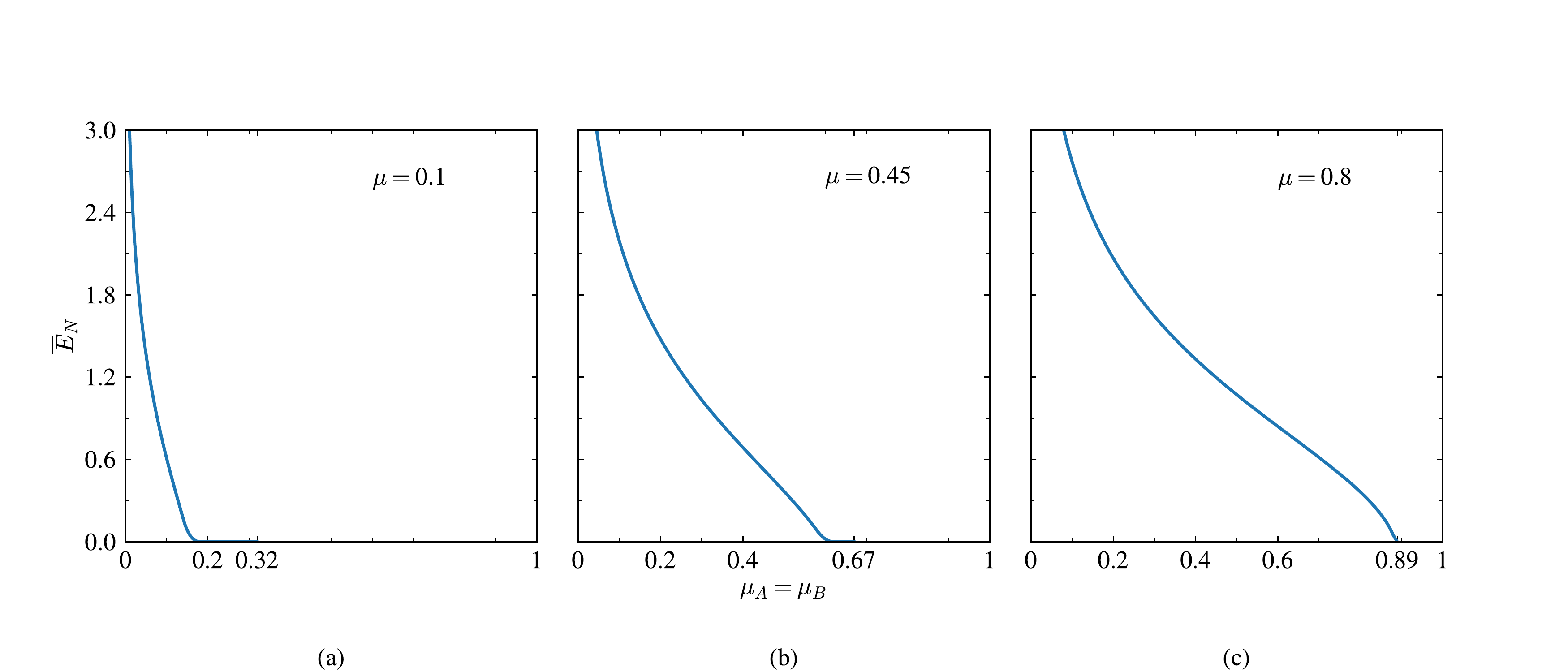}
\caption{{Average entanglement along a cut where $\mu_A=\mu_B$ 
for a) $\mu=0.1$, b) $\mu=0.45$ and c) $\mu=0.8$.
We see that the average entanglement decreases non-linearly
in all three cases, reaching zero in the separable domain and 
ceases to be well-defined in the unphysical domain $\mu_{A/B}>\sqrt{\mu}$.}}
\label{fig:cut_trough_regions}
\end{figure}

\subsection{Energy constrained typical quantum correlations}
{While the volumes of purity constrained subspaces of the last section were still infinite 
themselves (while their ratio was finite) we here consider finite volumes.}
A physically well motivated constraint that leads to a compact domain is to 
consider Gaussian state of fixed purity and fixed energy. An energy restriction for pure states has also been introduced in references~\cite{Lupo2012,Serafini2007}. The volume element we consider is
\begin{align}
 \diff V_{\mu,E}=\delta \Big(E-\frac{1}{2}\text{tr }\Sigma\Big)\frac{\sqrt{3}}{512}\,
 \frac{\mu^{7}}{\mu_A^3 \mu_B^3}\text{d}\mu_A\,\text{d}\mu_B\,\text{d}\Delta\,\text{d}\mu_1(S_A)\,\text{d}\mu_1(S_B)\,,
\end{align}
with $E$ from (\ref{eq:Energy}).
This is more involved than fixing marginal purities since the energy of a Gaussian state is not a local symplectic invariant, i.e. it depends not only on the marginal purities but also on the amount of local squeezing.
Of particular interest are again mean values of functions depending on the local symplectic invariants $\mu, \mu_A, \mu_B$ and $\Delta$, such as the logarithmic negativity and the steerability
\begin{equation}
\left\langle f(\mu, \mu_A, \mu_B, \Delta)\right\rangle |_{\mu,E} = \frac{\int\displaylimits_{\Sigma+i\Omega\geq 0}\mathrm{d}V_{\mu,E} f(\mu, \mu_A, \mu_B, \Delta)}{\int\displaylimits_{\Sigma+i\Omega\geq 0}\mathrm{d}V_{\mu,E}}\,.
\end{equation}
The relevant integrals can be simplified by carrying out the integration over the local symplectic groups, see appendix \ref{sec:deriv_Econstraint}. In particular we obtain
\begin{equation}
\begin{split}
  &\int\displaylimits_{\Sigma+\mathrm{i}\Omega\geq 0}\text{d}V_{\mu,E} f(\mu, \mu_A, \mu_B,\Delta) \\
 &=K''\hspace{-5pt} \int\displaylimits_{\sigma+\mathrm{i}\Omega\geq 0}\hspace{-10pt}\text{d}\mu_A\,\text{d}\mu_B\,\text{d}\Delta\frac{\mu^{7}}{\mu_A^2 \mu_B^2}
   \Big(E-\Big(\frac{1}{\mu_A}+\frac{1}{\mu_B}\Big)\Big)\Theta\Big(E-\Big(\frac{1}{\mu_A}+\frac{1}{\mu_B}\Big)\Big)f(\mu,\mu_A,\mu_B,\Delta)
\end{split}
\label{eq:fintegral}
\end{equation}
where $K''$ is a constant. To explicitly compute the integrals we
first analytically solve the $\Delta$-integral and then treat the
remaining two-dimensional integral over the marginal purities
numerically with an iterative and adaptive Monte Carlo method
\citep{Lepage1978, Lepage1980}. \\ In Fig. \ref{fig:proportions} the
proportion of entangled and steerable states, as well as typical
values for the logarithmic negativity and steerability, are displayed
for four different energies as functions of the purity. Each curve
starts at a different point which is the minimal possible purity for
the given energy.  Higher values of the purity allow for energy to be 
distributed to squeezing which can generate entanglement between the
two modes. As a result all curves are monotonically increasing. Since
all steerable states are entangled the proportion of steerable states
is always smaller than the proportion of entangled states.\\ The
results for pure states $\mu=1$ (big dots) are computed using the
invariant measure on the symplectic group, from
\cite{Lupo2012}. Almost all pure states are entangled and steerable,
thus the curves in Figs. \ref{fig:proportions} (a) and
\ref{fig:proportions} (c) reach one at $\mu=1$.
% N
% \begin{figure}[!htb]
% \centering
% \begin{subfigure}{0.50\textwidth}
% \flushright
% \includegraphics[width=\textwidth]{Plots/Physical.pdf}
% \caption{\label{fig:physical}}
% \end{subfigure}%
% \begin{subfigure}{0.50\textwidth}
% \flushright
% \includegraphics[width=\textwidth]{Plots/Coexistence.pdf}
% \caption{\label{fig:coexistence}}
% \end{subfigure}
% \caption{(a) Volume of physical states in arbitrary units and (b) proportion of coexistence region in the region of physical states.}
% \label{fig:plainvolumes}
% \end{figure}
% The total volume of the set of physical states shown in
% fig. \ref{fig:physical} increases with the energy.  \comment{Ideas
% what to describe/discuss:\\ With increasing energy the threshold as
% well as the peak move to lower purities. For large purities as well as
% for low energies the volume of physical states tends to
% zero. Comparing fig.\ \ref{fig:physical} and fig.\ \ref{fig:entangled}
% we see that first there are only separable states. In the coexistence
% region it is not possible to decide whether a state is entangled or
% not knowing only the purities. The proportion of this region in the
% physical states peaks for low purities with an increasing peak height
% for lower energies and tends to zero for purities close to one.}

\begin{figure}[!htb]
\includegraphics[width=\textwidth]{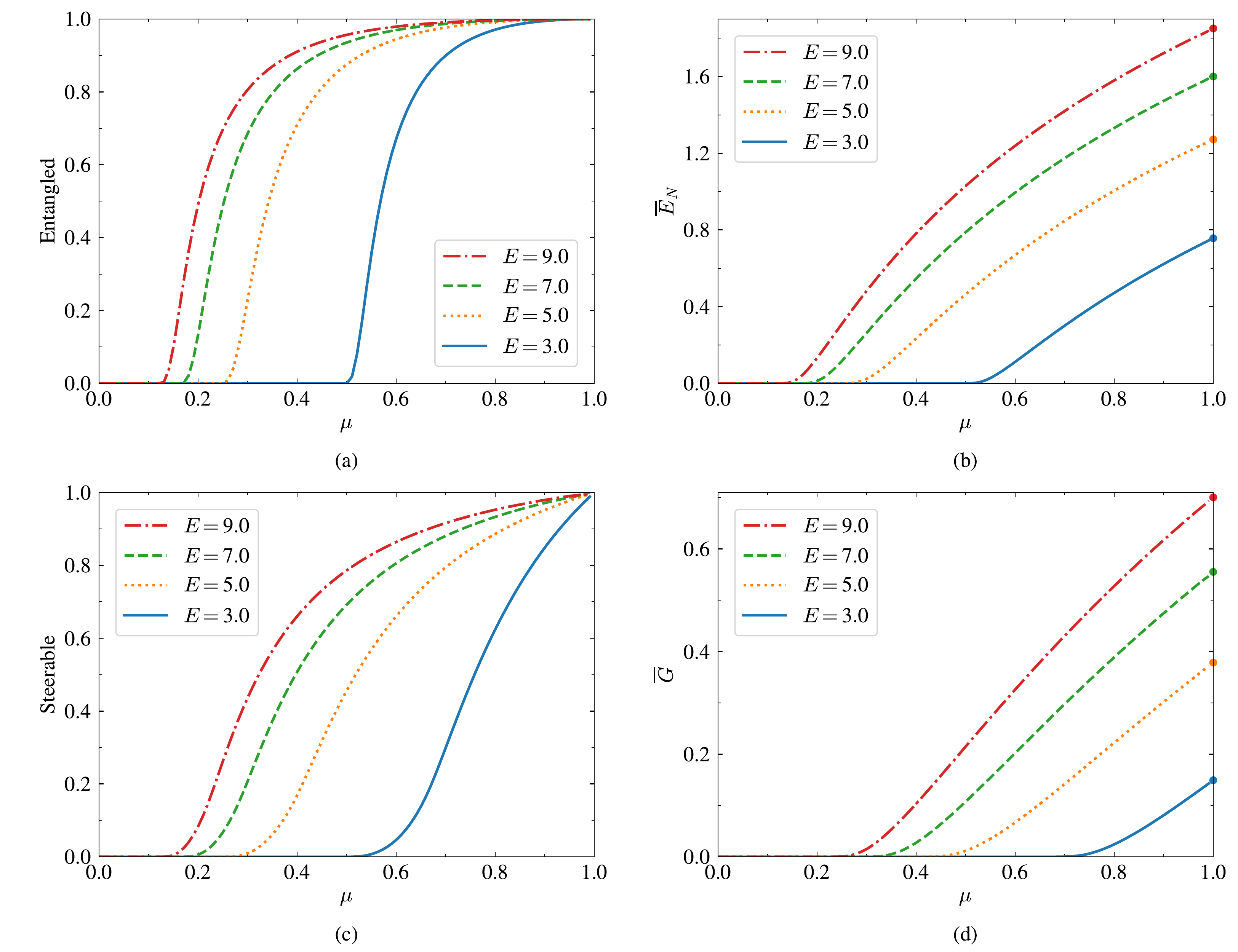}
\caption{(a) Proportion of entangled states (b) mean logarithmic
negativity (c) proportion of steerable states (d) mean
steerability. The dots at $\mu = 1$ in plots (b) and (d) show the
results computed with the measure for pure Gaussian states.}
\label{fig:proportions}
\end{figure}
% 
% \comment{Figure \ref{fig:proportions} shows the proportions of entangled and steerable states and the quantification thereof. The overall trend is the same. States appear above a threshold and for purities close to one the proportions go to one. It is noticeable that for all plots, the values of smaller energies are not higher than those of larger energies. It is also easy to see that the steerable states are a strict subset of the entangled states and the steerability never exceeds the entanglement. Due to the very small sampling domain for large purities, it is difficult to compute values for purities close one. In Figs. \ref{fig:entanglement} and \ref{fig:steerability} the dots mark denote values computed with the energy restricted invariant measure for pure states. Since these agree very well with our values close to the one, it can be assumed that the emerging behaviour will continue in the gap.\\
% }
% 

\section{Conclusions and outlook}\label{sec:concl}
In this article we have shown that three different 
unbiased measures for mixed Gaussian states are equivalent when 
constrained on the states with fixed purity. This result is 
somewhat surprising. The {rigorous} equivalence observed in 
the Gaussian case is at variance with numerical results obtained 
for finite dimensional systems in \cite{Zyczkowski1999}, where
different measures were close to each other but not equivalent.
We then proposed a unique unbiased measure for fixed purity Gaussian
states.

With this result a volume element suitable to compute typical correlation
properties of two mode ($1\times 1$) mixed Gaussian states is constructed. We first investigated
whether a typical state with given global and marginal purities is entangled or separable and then quantified the typical amount
of entanglement. In this situation a region of coexistence exists, where 
it is not possible to discriminate whether a
state is entangled or separable by purity measurements alone. Using 
the unique invariant measure we were able to compute  
the typical entanglement in the whole state space, allowing us to also characterize 
the coexistence region. 

A second way to resolve {the problems arising from integrating over a non-compact state 
space} is to consider compact subspaces by fixing the energy of the state. 
This is a physically well motivated restriction that has been suggested by others before. 
For high purities our results converge to the typical pure state values in the Haar invariant measure of the symplectic group.

{In the future, we will use our results to study 
generic properties of Gaussian channels via the
Choi-Jamiolkowski isomorphism~\cite{Holevo2011,Kiukas2017} 
and via probe states with limited resources~\cite{Monras2010} and compare the 
two approaches. Also our results could be used for channel discrimination
tasks that require optimization over the probe states \cite{Invernizzi2011}, or even to characterize the distinguishability of CV channels.
In the probe state approach, there is naturally a limited amount of resources 
available for the experimenter, such as states with limited energy. This underlines the usefulness of the energy constraint used in this work.}
{On the other hand, the constraints on the global and local purities, even if the volumes are infinite, lead
to rather generic and elegant results for typical values of symplectic invariants.}

\begin{acknowledgments}
The authors would like to thank Gerardo Adesso, Antti Karlsson, Simon Milz, H. Chau Nguyen, Roope Uola and Karol $\dot{\mathrm{Z}}$yczkowski for insightful discussions.
V.L. acknowledges support from the International Max Planck Research School (IMPRS) of MPIPKS Dresden.
\end{acknowledgments}

\appendix
\section{Volume element in Fisher-Rao metric}\label{sec:deriv_FR}
We use the strategy of \cite{Link2015}
to derive the Fisher-Rao volume element explicitly for the symplectic eigenvalue decomposition $\Sigma=S^TDS$. The infinitesimal shift in $\Sigma$ can be written as $\diff\Sigma =S\transp(\diff D+\diff H D+D\diff H)S$, where
$H$ is a special Hamiltonian matrix as defined in \cite{Link2015} 
\begin{align}
 H=\begin{pmatrix}
    H^{(11)}&\hdots & H^{(1N)}\\
    \vdots&\ddots  &\vdots\\
    H^{(N1)}&\hdots &H^{(NN)}
   \end{pmatrix}\,,\qquad
   H^{(ij)}=\begin{pmatrix}
           X_{ij}& Y_{ij}\\
           Z_{ij}&-X_{ji}
          \end{pmatrix}\,,
\end{align}
with $N \times N$ matrices $X$, $Y=Y^T,\,
 Z=Z^T$ and $Z$ having vanishing diagonal $Z_{ii}=0$ \footnote{Note the difference in ordering of position and momentum compared to the reference}. 
The line element of the Fisher Rao metric for Gaussian states is
\begin{align}
 \diff s_{\mathrm{FR}}^2=\frac{1}{2}\text{tr}\big((\Sigma^{-1}\text{d} \Sigma)^2\big)=\,\frac{1}{2}\text{tr}\big((D^{-1}\text{d} D)^2\big)+\text{tr}\big((\text{d}H)^2\big)
  +\text{tr}(D^{-1}\text{d}H\transp D\text{d}H)\,.
\end{align}
The single terms give explicitly
\begin{equation}
\text{tr}\big((D^{-1}\text{d}D)^2\big)=2\sum_{i=1}^N\frac{\text{d}\nu_i^2}{\nu_i^2}\,,
\end{equation}
\begin{equation}
\begin{split}
 \text{tr}\big((\text{d}H)^2\big)
 %&=2\text{tr}\big((\text{d}A)^2\big)+2\text{tr}(\text{d}B\text{d}C)\\
 &=2\sum_{i,j=1}^N\text{d}X_{ij}\text{d}X_{ji}+2\sum_{i>j=1}^N(\text{d}Y_{ij}\text{d}Z_{ij}+\text{d}Z_{ij}\text{d}Y_{ij})\,,
\end{split}
\end{equation}
 \begin{equation}
\begin{split}
 \text{tr}(\text{d}HD^{-1}\text{d}H\transp D)
 %&= \text{tr}(\text{d}A\mathcal{N}^{-1}\text{d}A\transp \mathcal{N})+\text{tr}(\text{d}A\transp \mathcal{N}^{-1}\text{d}A\mathcal{N})+\\
 %&\quad+\text{tr}(\text{d}B\mathcal{N}^{-1}\text{d}B\mathcal{N})+\text{tr}(\text{d}C\mathcal{N}^{-1}\text{d}C\mathcal{N})\\
 &=\sum_{i,j=1}^N\text{d}X_{ij}^2\Big(\frac{\nu_i}{\nu_j}+\frac{\nu_j}{\nu_i}\Big)+\sum_{i>j=1}^N(\text{d}Y_{ij}^2+\text{d}Z_{ij}^2)\Big(\frac{\nu_i}{\nu_j}+\frac{\nu_j}{\nu_i}\Big)+\\
 &\quad+\sum_{i=1}^N\text{d}Y_{ii}^2\,.
\end{split}
 \end{equation}
Overall, the distance element is
\begin{equation}
\begin{split}
\text{d}s_{\mathrm{FR}}^2=&
2\begin{pmatrix}
\text{d}\nu_1\\
\vdots\\
\text{d}\nu_N
\end{pmatrix}\transp
\begin{pmatrix}
    \nu_1^{-2}& &0\\
     &\ddots & \\
     0& & \nu_N^{-2}
   \end{pmatrix}
%\mathrm{diag}(\nu_1^{-2},...,\nu_N^{-2})
\begin{pmatrix}
\text{d}\nu_1\\
\vdots\\
\text{d}\nu_N
\end{pmatrix}
+\begin{pmatrix}
\text{d}X_{11}\\
\vdots\\
\text{d}X_{NN}
\end{pmatrix}\transp
\,\boldsymbol{1}
\begin{pmatrix}
\text{d}X_{11}\\
\vdots\\
\text{d}X_{NN}
\end{pmatrix}+\\
&+\sum_{i>j=1}^N
\begin{pmatrix}
  \text{d}X_{ij}\\
  \text{d}X_{ji}
 \end{pmatrix}\transp
 \begin{pmatrix}
  \frac{\nu_i}{\nu_j}+\frac{\nu_j}{\nu_i}&2\\
  2&\frac{\nu_i}{\nu_j}+\frac{\nu_j}{\nu_i}
 \end{pmatrix}
\begin{pmatrix}
  \text{d}X_{ij}\\
  \text{d}X_{ji}
 \end{pmatrix}
+\begin{pmatrix}
\text{d}Y_{11}\\
\vdots\\
\text{d}Y_{NN}
\end{pmatrix}\transp
\boldsymbol{1}
\begin{pmatrix}
\text{d}Y_{11}\\
\vdots\\
\text{d}Y_{NN}
\end{pmatrix}+\\
&+\sum_{i>j=1}^N
\begin{pmatrix}
  \text{d}Y_{ij}\\
  \text{d}Z_{ij}
 \end{pmatrix}\transp
 \begin{pmatrix}
  \frac{\nu_i}{\nu_j}+\frac{\nu_j}{\nu_i}&2\\
  2&\frac{\nu_i}{\nu_j}+\frac{\nu_j}{\nu_i}
 \end{pmatrix}
\begin{pmatrix}
  \text{d}Y_{ij}\\
  \text{d}Z_{ij}
 \end{pmatrix}\,.
\end{split}
\end{equation}
One can read the explicit form of the metric tensor. The measure $\sqrt{\det g}$ turns out to be
\begin{equation}
\begin{split}
  \sqrt{\det g}&=\sqrt{\det \big(\frac{\delta_{ij}}{\nu_i\nu_j}\big)}
 \prod\limits_{l>m=1}^N\Big(\big(\frac{\nu_i}{\nu_j}+\frac{\nu_j}{\nu_i}\big)^2-4\Big)\\
 &=\sqrt{\frac{1}{\prod_{k=1}^N\nu_k^2}\det \big(\delta_{ij}\big)} \Big(\prod_{k=1}^N\frac{1}{\nu_k}\Big)^{2(N-1)}\prod\limits_{l>m=1}^N(\nu_l^2-\nu_m^2)^2\\
 &=\Big(\prod_{k=1}^N\frac{1}{\nu_k}\Big)^{2N-1}
 \prod\limits_{l>m=1}^N(\nu_l^2-\nu_m^2)^2\,.
 \label{eq:explmeasure2}
\end{split}
\end{equation}

\section{Two-mode Hilbert-Schmidt volume element}\label{sec:two-mode-hilbert}
% The Hilbert-Schmidt line element is given by \cite{Link2015}
% \begin{equation}
%  \text{d}s_{\mathrm{HS}}^2
% =\frac{1}{16\sqrt{\det\Sigma}}\left(\tr{\Sigma^{-1}\diff\Sigma}{}^2
%      +2\tr{(\Sigma^{-1}\diff\Sigma)^2}{}\right)\,.
% \end{equation}
Any two mode covariance matrix can be written in the standard form \cite{Adesso2007}
\begin{align}
  \Sigma = S^T\sigma S\,, \qquad \sigma=\begin{pmatrix}a &0 & c_+& 0\\ 0&a&0&c_-\\ c_+&0&b&0\\
                                         0&c_-&0&b
                                        \end{pmatrix}  \,,
  \label{eq:stdform}
\end{align}
with the local symplectic transformation $S=S_A\oplus S_B,\textrm{and}\, S_A,S_B\in\text{Sp}(2)$.
Thus we may write
\begin{equation}
 \text{d}\Sigma=S\transp (\text{d}\sigma+\text{d}H\transp \sigma+\sigma\text{d}H  )S\,,
\end{equation}
where $\id+\text{d}H$ is {an} infinitesimal local symplectic transformation, i.e. $H$ is a Hamiltonian matrix with
\begin{equation}
 H=\begin{pmatrix}
    x_{A} & y_A\\
    z_A & -x_A
   \end{pmatrix}\oplus\begin{pmatrix}
    x_B & y_B\\
    z_B & -x_B
   \end{pmatrix}\,.
\end{equation}
We can now insert this parametrization in the expression for the Hilbert-Schmidt line element (\ref{eq:HS_dist})  to compute the metric tensor. One obtains the volume element
\begin{equation}
\diff V_{\mathrm{HS}}=
  \frac{ \sqrt{3} }{256}\frac{a^2 b^2 \left(c_+^2-c_-^2\right)}
    {\left(c_+^2-a b\right)^{5} \left(c_-^2-a b\right)^{5}}
    \diff a\,\diff b\,\diff c_+\,\diff c_-\, \diff x_A\,\diff x_B\,
    \diff y_A\,\diff y_B\,\diff z_A\,\diff z_B\,.
\end{equation}
The invariant measure on the symplectic group is given by $\diff \mu_1(S_{A/B})=\diff x_{A/B}\,\diff y_{A/B}\, \diff z_{A/B}$. The parameters $a, b, c_\pm$ can be expressed by the local purities $\mu_A, \mu_B$ and the symplectic invariants $\mu$ and $\Delta$ \cite{Adesso2004}
\begin{equation}
 \mu_A=1/a\,,\quad \mu_B=1/b\,,\quad \mu=\sqrt{(ab-c_+^2)(ab-c_-^2)}\,,\quad \Delta=a^2+b^2+2c_+c_-
\end{equation}
Switching to these coordinates gives
\begin{align}
  \diff V_{\mathrm{HS}}=& \frac{\sqrt{3}}{512}\frac{\mu^7}{\mu_A^3\mu_B^3}\diff\mu_A\,\diff\mu_B\,\diff\mu\,\diff\Delta\,\diff\mu_1(S_A)\diff\mu_1(S_B).
  \label{eq:HSmuVolumeelement}
\end{align}

%\section{Hilbert-Schmidt measure in purity-seralian coordinates\label{sec:deriv_HS_purity_seralian}}
%Any two mode covariance matrix can be written in the standard form \cite{Adesso2007}
%\begin{align}
%  \Sigma = S\transp \sigma S, && S=S_A\oplus S_B, && S_A,S_B\in\mathrm{Sp}(2),&&  \sigma=\matt{a\id & C \\ C & b\id}, && C=\matt{c_+& 0 \\ 0 & c_-},
%\end{align}
%where $a, b, c_\pm$ are functions of the invariants $\mu, \mu_A, \mu_B$ and $\Delta$ \cite{Adesso2004}. In these variables the Hilbert-Schmidt measure becomes
%\begin{align}
%  \diff V_{HS}=& 4\sqrt{3} a^2 b^2(c_+^2-c_-^2)(16(ab-c_+^2)(ab-c_-^2))^{-5}\diff a\,\diff b\,\diff c_+\,\diff c_-\,\diff\mu_1(S_1)\,\diff\mu_1(S_2).
%\end{align}

\section{Energy constraint}\label{sec:deriv_Econstraint}
The single mode symplectic operation $S_{A}\in \mathrm{Sp}(2)$ can be written
as
\begin{align}
  S_{A} = O'WO, && O,O' \in \mathrm{SO}(2), && W=\matt{w & 0 \\ 0 & 1/w}, && w\geq 1.
\end{align}
Using this, the invariant measure over the symplectic group can be
decomposed in a compact part corresponding to rotations and a non-compact
part corresponding to single mode squeezing \cite{Lupo2012}
\begin{align}
\diff\mu_1(S_{A})=K\diff\lambda_{A}\diff\mu(O)\diff\mu(O'),&& \lambda_{A}=\frac{1}{2}(w^2+1/w^2),
\end{align}
where $\diff\mu(O)$ is an invariant measure over $\mathrm{SO}(2)$ and $K$ is a normalization
factor. The energy of a single mode covariance matrix, written as $\Sigma_{A/B}=\mu_{A}^{-1}S_{A}\transp S_{A}$ is
then
\begin{align}
  E_{A}=\frac{1}{2}\text{tr }\Sigma_{A}=\frac{\lambda_{A}}{\mu_{A}}.
\end{align}
The energy of the two mode CM $\Sigma$ is given by the sum of the energies of the single mode subsystems $A$ and $B$
\begin{equation}
 E=\frac{1}{2}\text{tr }\Sigma=\frac{\lambda_A}{\mu_A}+\frac{\lambda_B}{\mu_B}
\end{equation}
We can carry out the integrals over the local symplectic groups respecting an energy constraint
\begin{align}
 \int\displaylimits_{\Sigma+i\Omega\geq 0}&\diff V_{\mu,E} f(\mu, \mu_A, \mu_B,\Delta)
 =K'\int\displaylimits_{\sigma+i\Omega\geq 0}\diff\mu_A\diff\mu_B\diff \Delta\,\frac{\mu^{7}}{\mu_A^3 \mu_B^3}f(\mu,\mu_A,\mu_B,\Delta)\times\notag\\
  &\times\int_1^\infty \text{d}\lambda_A \text{d}\lambda_B \delta\Big(E-\Big(\frac{\lambda_A}{\mu_A}+\frac{\lambda_B}{\mu_B}\Big)\Big)
    \int \text{d}\mu(O_A)\text{d}\mu(O'_A)\text{d}\mu(O_B)\text{d}\mu(O'_B)\notag\\
 &=K'' \int\displaylimits_{\sigma+i\Omega\geq 0}\text{d}\mu_A\text{d}\mu_B\text{d}\Delta\,\frac{\mu^{7}}{\mu_A^3 \mu_B^3}
   f(\mu,\mu_A,\mu_B,\Delta)
   \int\!\!\!\int_1^\infty \text{d}\lambda_A \text{d}\lambda_B \delta\Big(E-\Big(\frac{\lambda_A}{\mu_A}+\frac{\lambda_B}{\mu_B}\Big)\Big)
\end{align}
Computing the $\lambda$-integrals over the delta function gives %(use $E'=\frac{\lambda_A}{\mu_A}+\frac{\lambda_B}{\mu_B}$ and $\varepsilon=\frac{\lambda_A}{\mu_A}-\frac{\lambda_B}{\mu_B}$)
% \begin{align}
%   &\int\!\!\!\int_{-\infty}^\infty \text{d}\lambda_A \text{d}\lambda_B \delta\Big(E-\Big(\frac{\lambda_A}{\mu_A}+\frac{\lambda_B}{\mu_B}\Big)\Big)
%     \Theta(\lambda_A-1)\Theta(\lambda_B-1)\notag \\
%   &=\int_{-\infty}^\infty  \text{d}E'\text{d}\varepsilon\, \delta(E-E')
% \Theta\Big(\frac{\mu_A}{2}(E'+\varepsilon)-1\Big)\Theta\Big(\frac{\mu_B}{2}(E'-\varepsilon)-1\Big)\frac{1}{2}\mu_A\mu_B\notag\\
% &=\mu_A\mu_B\Big(E-\Big(\frac{1}{\mu_A}+\frac{1}{\mu_B}\Big)\Big)\Theta\Big(E-\Big(\frac{1}{\mu_A}+\frac{1}{\mu_B}\Big)\Big)
% \end{align}
% and therefore
\begin{align}
 &\int\displaylimits_{\Sigma+i\Omega\geq 0}\text{d}V_{\mu,E} f(\mu, \mu_A, \mu_B,\Delta) \\
 &=K''\hspace{-5pt} \int\displaylimits_{\sigma+i\Omega\geq 0}\hspace{-10pt}\text{d}\mu_A\text{d}\mu_B\text{d}\Delta\frac{\mu^{7}}{\mu_A^2 \mu_B^2}
\Big(E-\Big(\frac{1}{\mu_A}+\frac{1}{\mu_B}\Big)\Big)\Theta\Big(E-\Big(\frac{1}{\mu_A}+\frac{1}{\mu_B}\Big)\Big)f(\mu,\mu_A,\mu_B,\Delta)
\end{align}

\bibliography{TypicalGaussian.bib}
\end{document}